\documentclass[11pt]{article} 
\usepackage{color}
\usepackage{amsmath,amsfonts,amsthm,amssymb}
\usepackage{times}
\usepackage{eucal}
\usepackage{fullpage}

\DeclareMathOperator{\Exp}{Exp}
\DeclareMathOperator{\Var}{Var}

\numberwithin{equation}{section}

\newtheorem{theorem}{Theorem}
\newtheorem{lemma}[theorem]{Lemma}

\newtheorem{claim}[theorem]{Claim}
\newtheorem{corollary}[theorem]{Corollary}
\newtheorem{definition}{Definition}
\DeclareMathAlphabet{\varmathbb}{U}{bbold}{m}{n}
\newcommand{\one}{\varmathbb 1}
\renewcommand{\vec}[1]{\mathbf{#1}}

\newcommand{\remove}[1]{}

\newcommand{\C}{\mathbb{C}}

\newcommand{\Z}{\mathbb{Z}}

\newcommand{\ket}[1]{\left| #1 \right\rangle}
\newcommand{\bra}[1]{\left\langle #1 \right|}

\newcommand{\CG}{\C[G]}

\newcommand{\rank}{\textbf{rk}\;}
\newcommand{\norm}[1]{\left\| #1 \right\|}
\newcommand{\abs}[1]{\left| #1 \right|}
\newcommand{\U}{\textsf{U}}

\newcommand{\inner}[2]{\left\langle #1, #2 \right\rangle}

\newcommand{\poly}{{\rm poly}}

\newcommand{\vb}{\vec{b}}

\newcommand{\vv}{\vec{v}}

\newcommand{\vrho}{{\boldsymbol{\rho}}}

\newcommand{\wg}{\widehat{G}}
\newcommand{\wk}{\widehat{K}}
\newcommand{\Ind}{\text{Ind}}
\newcommand{\isotype}{\mathfrak{I}}
\newcommand{\uniform}{\mathcal{U}}
\newcommand{\average}{\mathcal{A}}
\newcommand{\planch}{\mathcal{P}}
\newcommand{\measured}{\mathcal{H}}
\newcommand{\trs}{\tau_{\{\rho,\sigma\} }}

\title{Tight Results on Multiregister Fourier Sampling:\\ Quantum
  Measurements for Graph Isomorphism Require Entanglement}

\author{Cristopher Moore \\
  \textsf{moore@cs.unm.edu}\\
  Department of Computer Science\\
  University of New Mexico
  \and
  Alexander Russell\\
  \textsf{acr@cse.uconn.edu}\\
  Department of Computer Science and Engineering\\
  University of Connecticut}

\begin{document}
\begin{titlepage}
  \maketitle 

  
  \begin{abstract}
    We establish a general method for proving bounds on the
    information that can be extracted via arbitrary entangled
    measurements on tensor products of hidden subgroup coset states.
    When applied to the symmetric group, the method yields an $\Omega(n
    \log n)$ lower bound on the number of coset states over which we
    must perform an entangled measurement in order to obtain
    non-negligible information about a hidden involution.  These results
    are tight to within a multiplicative constant and apply, in
    particular, to the case relevant for the Graph Isomorphism
    problem.
    
    Part of our proof was obtained after learning from Hallgren,
    R\"otteler, and Sen that they had obtained similar results.
  \end{abstract}
\end{titlepage}

\section{Introduction: the hidden subgroup problem}

Many problems of interest in quantum computing can be
reduced to an instance of the \emph{Hidden Subgroup Problem}
(HSP). This is the problem of determining a subgroup $H$ of a group $G$
given oracle access to a function $f: G \to S$ with the property that
$
f(g) = f(hg) \Leftrightarrow h \in H 
$.  
Equivalently, $f$ is constant on the cosets of $H$ and takes
distinct values on distinct cosets.

All known efficient solutions to the problem rely on the
\emph{standard method} 
\cite{BernsteinV93}, in which we prepare a uniform 
superposition over the elements of $G$ and measure the value of the oracle 
on this superposition.  This yields a uniform superposition over a
uniformly random left coset, $\ket{cH} = (1/\sqrt{|H|}) \sum_{h \in H} \ket{ch}$, 
or equivalently a mixed state,
$
\rho_H = (1/|G|) 
\sum_{c \in G} \ket{cH}\bra{cH} 
$.
The question is how much information about the subgroup $H$ can be gained by measuring this state.  {\em Fourier sampling} measures $\rho_H$ according to the Fourier basis, i.e., according to the irreducible representations of $G$; as we discuss below, the optimal measurement is always of this type.

\paragraph{History of the Hidden Subgroup Problem.}
Both Simon's and Shor's seminal algorithms rely on the standard method
over an abelian group. In Simon's problem~\cite{Simon97}, $G=\Z_2^n$
and $f$ is an oracle such that, for some $y$, $f(x) = f(x+y)$ for all
$x$; in this case $H=\{0,y\}$ and we wish to identify $y$.  In Shor's
factoring algorithm~\cite{Shor97} $G$ is (essentially) the group $\Z_n^*$ where $n$
is the number we wish to factor, $f(x) = r^x \bmod n$ for a random $r
< n$, and $H$ is the subgroup of $\Z_n^*$ whose index is the
multiplicative order of $r$.  
\remove{
(Note that in Shor's algorithm, since
$|\Z_n^*|$ is unknown, the Fourier transform is performed over $\Z_q$
for some $q=\poly(n)$; see \cite{Shor97} or \cite{HalesH99,HalesH00}.)
}

\remove{
In abelian groups, Fourier sampling consists of observing one of the frequencies, 
or {\em characters}, of the group.  It is not hard to see that a polynomial
number (i.e., polynomial in $\log |G|$) of experiments of this type
determine $H$.  In essence, each experiment yields a random element of
the dual space $H^\perp$ perpendicular to $H$'s characteristic function,
and as soon as these elements span $H^\perp$ they, in particular,
determine $H$.
}

While the \emph{nonabelian} hidden subgroup problem appears to be much
more difficult, it has very attractive applications.  In particular,
solving the HSP for the symmetric group $S_n$ would provide an
efficient quantum algorithm for the Graph Automorphism and Graph
Isomorphism problems (see e.g.\ Jozsa~\cite{Jozsa00} for a review).
Another important motivation is the relationship between the HSP over
the dihedral group with hidden shift problems~\cite{vanDamHI03} and
cryptographically important cases of the Shortest Lattice Vector
problem~\cite{Regev}.

So far, algorithms for the HSP are only known for a few families of
nonabelian
groups~\cite{RoettelerB98,IvanyosMS01,FriedlIMSS02,MooreRRS04,InuiLeGall,BaconCvD2}.
\remove{, including wreath products $\Z_2^k\; \wr\;
  \Z_2$~\cite{RoettelerB98}; more generally, semidirect products $K
  \ltimes \Z_2^k$ where $K$ is of polynomial size, and groups whose
  commutator subgroup is of polynomial size~\cite{IvanyosMS01};
  ``smoothly solvable'' groups~\cite{FriedlIMSS02}; and some
  semidirect products of cyclic groups~\cite{InuiLeGall}.  } Ettinger
and H{\o}yer~\cite{EttingerH98} provided another type of result (see
also~\cite{RadhakrishnanRS}) by showing that Fourier sampling can
solve the HSP for the dihedral groups $D_n$ in an
\emph{information-theoretic} sense.  That is, a polynomial number of
experiments gives enough information to reconstruct the subgroup,
though it is unfortunately unknown how to determine $H$ from this
information in polynomial time.

To discuss Fourier sampling for a nonabelian group $G$, one needs to
consider \remove{ develop the Fourier transform over $G$. For abelian
  groups, the Fourier basis functions are homomorphisms $\phi:G \to \C$
  such as the familiar exponential function $\phi_k(x) = e^{2\pi i kx/n}$
  for the cyclic group $\Z_n$.  In the nonabelian case, there are not
  enough such homomorphisms to span the space of all $\C$-valued
  functions on $G$; to complete the picture, one introduces
}\emph{representations} of the group, namely homomorphisms $\rho:G \to
\U(V)$ where $\U(V)$ is the group of unitary matrices acting on some
$\C$-vector space $V$ of dimension $d_\rho$. It suffices to consider
\emph{irreducible} representations, namely those for which no
nontrivial subspace of $V$ is fixed by the various operators $\rho(g)$.
Once a basis for each irreducible $\rho$ is chosen, the matrix elements
$\rho_{ij}$ provide an orthogonal basis for the vector space of all
$\C$-valued functions on $G$.
  The quantum Fourier transform then consists of transforming
  (unit-length) vectors in $\CG = \{ \sum_{g \in G} \alpha_g \ket{g} \mid \alpha_g
  \in \C\}$ from the basis $\{ \ket{g} \mid g \in G \}$ to the basis $\{
  \ket{\rho,i,j} \}$ where $\rho$ is the name of an irreducible
  representation and $1 \leq i, j \leq d_\rho$ index a row and column (in a
  chosen basis for $V$).  Indeed, this transformation can be carried
  out efficiently for a wide variety of
  groups~\cite{Beals97,Hoyer97,MooreRR04}. 
  \remove{Note, however, that a
  nonabelian group $G$ does not distinguish any specific basis for its
  irreducible representations which necessitates a rather dramatic
  choice on the part of the transform designer. Indeed, careful basis
  selection appears to be critical for obtaining efficient Fourier
  transforms for the groups mentioned above.
  }

  A basic question concerning the hidden subgroup problem is whether
  there is always a basis for the representations of $G$ such that
  measuring in this basis provides enough information to determine the
  subgroup $H$. This framework is known as \emph{strong Fourier
    sampling}.  In~\cite{MooreRS}, Moore, Russell and Schulman
  answered this question in the negative, showing that subgroups of
  $S_n$ relevant to Graph Isomorphism cannot be determined by this
  process; more generally, they showed that no subexponential number
  of positive operator-valued measurements (POVMs) of individual coset
  states suffices.  \remove{ We emphasize that this result includes
    the most important special cases of the nonabelian HSP, as they
    are those to which Graph Isomorphism naturally reduces.  }
  \remove{ As in~\cite{HallgrenRT00} we shall focus on order-2
    subgroups of the form $\{1,m\}$, where $m$ is an involution
    consisting of $n/2$ disjoint transpositions; then if we fix two
    rigid connected graphs of size $n$ and consider permutations of
    their disjoint union, then the hidden subgroup is of this form if
    the graphs are isomorphic and trivial if they are not.  }

\newpage
  The next logical step is to consider \emph{multi-register}
  algorithms, in which we prepare multiple coset states and subject
  them to \emph{entangled} measurements.  Ettinger, H{\o}yer and
  Knill~\cite{EttingerHK04} showed that the HSP on arbitrary groups
  can be solved information-theoretically with a polynomial number of
  registers, and the authors of this article have shown how to carry
  out such a measurement for the case relevant to Graph Isomorphism in
  the Fourier basis~\cite{MooreR:banff}.  For the dihedral group
  $D_n$, Ip~\cite{Ip} showed that the optimal measurement for two
  registers is entangled, and Kuperberg~\cite{Kuperberg03} devised a
  subexponential ($2^{O(\sqrt{\log n})}$) algorithm that works by
  performing entangled measurements on two registers at a time.
  Bacon, Childs, and van Dam~\cite{BaconCvD1,BaconCvD2} determined the
  optimal multiregister measurement for certain metabelian groups, and
  use this to devise the first efficient multiregister algorithms.
  The present authors have generalized these optimality results to the
  case where $H$ and $G$ form a Gel'fand pair~\cite{MooreR:gelfand}.

  \paragraph{Our contribution.}
  Whether a similar approach can be applied to the symmetric group,
  offering an efficient algorithm for Graph Isomorphism, is the
  principal open question in this area.  Here we establish a general
  method for bounding the information that can be extracted by
  arbitrary entangled measurements on tensor products of coset states.
  These bounds give rise to the following theorem:

\begin{theorem}
\label{thm:main}
Suppose we are given the coset state $\rho_H^{\otimes k}$ on $k$
registers for the hidden subgroup $H=\{1,m\}$ where $m$ is chosen
uniformly at random from a conjugacy class $M$ of involutions.  Given
that we observe the representation $\vrho = \rho_1 \otimes \cdots
\otimes \rho_k$, let $B$ be a basis for $\vrho$, let
$\measured_m(\vb)$ be the probability that we observe the vector $\vb
\in B$, and let $\uniform$ be the uniform distribution on $B$.  Then
there is a constant $C > 0$ such that, if $k < C n \log_2 n$, with
probability $1-n^{-\Omega(n)}$ in $m$ and $\vrho$, we have 
\[
\norm{
  \measured_m - \uniform }_1 = n^{-\Omega(n)}\enspace.
\]
\end{theorem}
\noindent
Thus, unless $k=\Omega(n \log n)$, it takes a superpolynomial number of
experiments to distinguish the different subgroups $H=\{1,m\}$ from each
other, or from the trivial subgroup, for which the observed
distribution is uniform.  Along with the fact that $O(n \log n)$
registers suffice~\cite{EttingerHK04,MooreR:banff}, this shows that
entangled measurements over $\Theta(n \log n)$ registers are both necessary
and sufficient.

Note that this result is much stronger than the claim that the total
query complexity of this case of the Hidden Subgroup Problem is $\Theta(n
\log n)$ (where each query consists of generating a coset state);
indeed, one can immediately obtain $\Omega(n)$ lower bounds on
the query complexity of determining an involution $m$ by embedding
$\Z_2^{n}$ into $S_{2n}$. In fact, these bounds can be obtained even
without the assumption that each query generates a coset
state~\cite{Koiran05}. The query complexity of the decision problem
of whether $H$ is of the form $\{1,m\}$ or is trivial was recently shown
to be $\Omega(n)$ in a natural hidden shift model~\cite{ChildsW}.

Such query complexity lower bounds, however, do not preclude the
possibility of using multiple independent applications of
(single-register) Fourier sampling to solve the problem; for instance,
in the dihedral group, each such measurement yields a constant amount
of information~\cite{EttingerH98}.  In contrast, the result proved
here shows that in order to gain non-negligible information about the
hidden subgroup, and thus about whether the two graphs are isomorphic
or not, one must measure $O(\log |G|)$ registers simultaneously in an
entangled basis.  This greatly restricts the set of possible quantum
algorithms for Graph Isomorphism.

\remove{
Detailed proofs and further discussion of the
results in the extended abstract can be found in the authors' quant-ph
postings~\cite[quant-ph/0501066]{prelim:2} and~\cite[quant-ph/0510233]{prelim:multi}.
}

\paragraph{Remark.} A preliminary version of this paper appeared
in~\cite{prelim:2} where we developed a general framework for bounding the 
available information in the multiregister case, including Lemmas 2--5, 
and showed that entangled measurements over two registers are insufficient. 
The proof of Lemma~\ref{lemma:projector-bounds} below, on which
Theorem~\ref{thm:main} depends, was obtained after learning from
Hallgren, R\"{o}tteler, and Sen that they had obtained results similar to 
Theorem~\ref{thm:main} by building on the machinery of~\cite{prelim:2}.

\section{The structure of the optimal measurement}

We focus on the special case of the hidden subgroup problem called the
\emph{hidden conjugate problem} in~\cite{MooreRRS04}.  Here there is a
(non-normal) subgroup $H$, and we are promised that the hidden
subgroup is one of its conjugates, $H^g = g^{-1} H g$ for some $g \in
G$; the goal is to determine which.

The most general possible measurement in quantum mechanics is a
positive operator-valued measurement (POVM).  It is easy to see~\cite{MooreRS} 
that the optimal POVM for the Hidden Subgroup Problem on a
single coset state consists of measuring the name $\rho$ of the
irreducible representation, followed by a POVM on the vector space $V$
on which $\rho$ acts.  For simplicity, here we will restrict ourselves to 
von Neumann measurements, in which we measure the space on which $\rho$ acts
according to some orthonormal basis $B$.  As in~\cite{MooreRS,prelim:2} our results 
can easily be extended to arbitrary POVMs.
\remove{
this corresponds to measuring the row of $\rho$ in some orthonormal
basis; in general it consists of measuring according to some
over-complete basis, or \emph{frame}, $B=\{ \vb \}$ with positive real
weights $a_\vb$ that obeys the completeness condition
$\sum_\vb a_\vb \pi_{\vb} = \one$,
where $\pi_\vb$ denotes the projection onto 
$\vb$.  For simplicity, here we will assume that $B$ is an 
orthonormal basis, so that $a_\vb = 1$ for all $\vb \in B$; 
}

Under Fourier sampling, the probability we observe $\rho$, and the
conditional probability that we observe a given $\vb \in B$, are given
by
\begin{equation}
 \measured(\rho)  =  \frac{d_\rho |H|}{|G|} \,\rank \Pi_H 
 \label{eq:prho2} \qquad\text{and}\qquad 
 \measured(\rho,\vb) = \frac{\norm{\Pi_H \vb}^2}{\rank \Pi_H} 
\end{equation}
where $\Pi_H$ is the projection operator $1/|H| \sum_{h \in H} \rho(h)$.
When $H$ is nontrivial, the probability distribution over $B$ 
changes for a conjugate $H^g$ in the following way:
\[ \measured(\rho,\vb) = \frac{\norm{\Pi_H g \vb}^2}{\rank \Pi_H} \]
where we write $g \vb$ for $\rho(g) \vb$.  In contrast, if $H$ is the
trivial subgroup, $\Pi_H = \one_{d_\rho}$ and $\measured(\rho)$ is the
\emph{Plancherel distribution} $\planch(\rho) = d_\rho^2/|G|$, and
$\measured(\rho,\vb_j)=1/d_\rho$ is the uniform distribution.

\section{The expectation and variance of an involution projector}

\begin{definition}
  \label{def:isotypic}
  Let $\rho$ be a representation of a group $G$ acting on a space $V$
  and let $\sigma$ be an irreducible representation of $G$. We let
  $\isotype_\sigma^\rho$ denote the projection operator onto the
  \emph{$\sigma$-isotypic} subspace of $V$, the subspace spanned by
  all copies of $\sigma$ in $\rho$. We remark that this projection
  operator can written as the sum $\isotype_\sigma^\rho \vv =
  \frac{d_\sigma}{|G|} \sum_g \chi_\sigma^*(g) g\vv$,
  regardless of the structure of $\rho$. See, e.g., \cite{Serre77}.
\end{definition}

The following two lemmas are proved in~\cite{MooreRS}; we repeat them here for
convenience.

\begin{lemma}
\label{lem:exp}
Let $\rho$ be a representation of a group $G$ acting on a space $V$ and
let $\vb \in V$.  Let $m$ be chosen uniformly from a
conjugacy class $M$ of involutions.  If $\rho$ is irreducible, then
\[ \Exp_m \langle \vb, m \vb \rangle = \frac{\chi_\rho(M)}{d_\rho} \norm{\vb}^2 \enspace . \]
If $\rho$ is reducible, then 
\[ \Exp_m \langle \vb, m \vb \rangle 
= \sum_{\sigma \in \wg} \frac{\chi_\sigma(M)}{d_\sigma} \norm{\isotype_\sigma^\rho \vb}^2 \enspace . \]
\end{lemma}

\begin{lemma}
\label{lem:second}
Let $\rho$ be a representation of a group $G$ acting on a space $V$ and
let $\vb \in V$.  Let $m$ be chosen uniformly from a
conjugacy class $M$ of involutions.  Then
\[ \Exp_m \abs{\langle \vb, m \vb \rangle}^2 
= \sum_{\sigma \in \wg} 
\frac{\chi_\sigma(M)}{d_\sigma} \norm{\isotype_\sigma^{\rho \otimes \rho^*} (\vb \otimes \vb^*)}^2 
\enspace . \]
\end{lemma}

Now, given an involution $m$ and the hidden subgroup $H=\{1,m\}$, let
$\Pi_m = \Pi_H$ denote the projection operator given by $\Pi_m \vv =
(1/2) (\vv + m \vv)$.  Then the expectation and variance of
$\norm{\Pi_m \vb}^2$ are given by the following lemma, also
from~\cite{MooreRS}.

\begin{lemma}  
\label{lem:var}
Let $\rho$ be an irreducible representation acting on a space $V$ and
let $\vb \in V$.  Let $m$ be chosen uniformly from a
conjugacy class $M$ of involutions.  Then
\begin{eqnarray}
  \Exp_{m} \norm{\Pi_m \vb}^2
  & = & \frac{1}{2} \norm{\vb}^2 \left(1 + \frac{\chi_\rho(M)}{d_\rho} \right) 
  \label{eq:exp} \\
  \Var_{m} \norm{\Pi_m \vb}^2 
  & \leq & \frac{1}{4} 
 \sum_{\sigma \in \wg} \frac{\chi_\sigma(M)}{d_\sigma}
   \norm{\isotype^{\rho \otimes \rho^*}_\sigma(\vb \otimes \vb^*)}^2 
  \enspace.
  \label{eq:var} 
\end{eqnarray}
\end{lemma}

\noindent
Finally, we point out that since $\Exp_{m} \norm{\Pi_m \vb}^2 =
\norm{\vb}^2 \frac{\rank \Pi_m}{d_\rho}$ we have
\begin{equation}
\label{eq:rank}
\frac{\rank \Pi_m}{d_\rho} = \frac{1}{2}  \left(1 + \frac{\chi_\rho(M)}{d_\rho} \right)
\enspace .
\end{equation}

\section{Variance and decomposition for multiregister experiments}
\label{sec:multi}

We turn now to the multi-register case, where Steps 1, 2 and 3 are
carried out on $k$ independent registers.  This yields a state in
$\C[G^k]$, i.e., $\ket{c_1 H} \otimes \cdots \otimes \ket{c_k H}$ where the $c_i$ are
uniformly random coset representatives.  The symmetry argument
of~\cite{MooreRS} applies to each register, so that the optimal
measurement is consistent with first measuring the representation name
in each register.  However, the optimal measurement generally does not
consist of $k$ independent measurements on this tensor product state;
rather, it is entangled, consisting of measurement in a basis whose
basis vectors $\vb$ are not of the form $\vb_1 \otimes \cdots \otimes \vb_k$.

In this section, we extend the results of~\cite{MooreRS} to the case of 
multiple coset states in three steps.  
First, in Section~\ref{sec:multi-var}, we generalize the expressions of 
Lemma~\ref{lem:var} for the expectation and variance of the observed 
distribution to the multiregister case. 
In Sections~\ref{sec:exp} and~\ref{sec:var}, we bound the expectation 
and variance of the probability distribution, by controlling to what extent 
tensor product vectors project into ``bad'' low-dimensional representations with 
large normalized characters.  These bounds are far tighter than those 
in~\cite{MooreRS,prelim:2}, in which we pessimistically bounded these projections 
simply by estimating the multiplicity of bad representations.
Finally, in Section~\ref{sec:tvd}, we combine these bounds to bound the 
expectation over $\vrho$ of the total variation distance between the observed 
distribution and the uniform distribution.

\subsection{Variance for Fourier sampling product states}
\label{sec:multi-var}

We begin by generalizing Lemmas~1, 2, and 3
of~\cite{MooreRS} to the multi-register case.  The reasoning is analogous
to that of Section 4 of~\cite{MooreRS}; the principal difficulty
is notational, and we ask the reader to bear with us. 

We assume we have measured the representation name on each of the
registers, and that we are currently in an irreducible representation
of $G^k$ labeled by $\vrho = \rho_1 \otimes \cdots \otimes \rho_k$.  For a subset $S \subset [k]$,
let us introduce the shorthand $\rho_S = \otimes_{i \in S} \rho_i$ and $\rho_S \otimes \one =
\bigotimes_{i \in S} \rho_i \otimes \bigotimes_{i \in \overline{S}} \one$, operating in the natural
way on the vector space that supports $\vrho$.

Then given a subset $I \subseteq [k]$, we can separate this tensor product
into the registers inside $I$ and those outside, and then decompose
the product of those inside $I$ into irreducibles:
\[
\vrho = \rho_I \otimes \rho_{\overline{I}} = \left( \bigoplus_{\sigma \in \wg} a^I_\sigma \sigma \right) \otimes
\rho_{\overline{I}}
\]
where $a^I_\sigma$ is the multiplicity of $\sigma$ in $\rho_I$.  Now given an irreducible representation 
$\sigma$, we let $\Pi^I_\sigma = \isotype_\sigma^{\rho_I \otimes \one}$ denote the projection
operator onto the subspace acted on by $a^I_\sigma \sigma \otimes \rho_{\overline{I}}$.
In other words, $\Pi^I_\sigma$ projects the registers in $I$ onto the
subspaces isomorphic to $\sigma$, and leaves the registers outside $I$
untouched.  Note that in the case where $I$ is a singleton we have
$\isotype^{\rho_i \otimes \one}_{\rho_i} = \one$.

As before, the hidden subgroup is $H=\{1,m\}$ for an involution $m$
chosen at random from a conjugacy class $M$.  However, we now have, in
effect, the subgroup $H^k \subset G^k$, and summing over the elements of
$H^k$ gives the projection operator $\Pi_{H^k} = \Pi_m^{\otimes k}$.  The
probability of observing a representation $\vrho$ under weak sampling
is thus
\[
\measured(\vrho) = \measured_M^{\otimes k}(\vrho) \triangleq \frac{d_\vrho |H|^k}{|G|^k} \bigl(\rank \Pi_H\bigr)^k\enspace.
\]
Conditioned upon observing $\vrho$, the probability we observe an
(arbitrarily entangled) basis vector $\vb \in \vrho$ is
\begin{equation}
  \label{eq:pbk}
  \measured(\vrho, \vb) =  \measured_m^{\otimes k}(\vrho, \vb) \triangleq
  \frac{\norm{\Pi_m^{\otimes k} \vb}^2}{\rank \Pi_m^{\otimes k}} \enspace.
\end{equation}
As indicated, we elide the superscripts and subscripts when they can
be inferred from context. We remark that the distribution
$\measured^{\otimes k}(\vrho)$ depends only on $M$ and can be written as a
product distribution: $\measured^{\otimes k}(\vrho) = \prod_i \measured^{\otimes
  1}(\rho_i)$. The distribution $\measured_m^{\otimes k}(\vrho, \vb)$, on the other hand, 
cannot in general be decomposed in this way as we consider arbitrarily entangled bases
of $\vrho$ as opposed to product bases.

When we calculate the expectation of this over $m$, we will find
ourselves summing the following quantity over the subsets $I \subseteq [k]$:
\begin{equation}
\label{eq:ei}
 E^I(\vb) 
 \triangleq \sum_{\sigma \in \wg} 
\frac{\chi_\sigma(M)}{d_\sigma} \norm{\Pi^I_\sigma \vb}^2 
\end{equation}
with $E^\emptyset(\vb) = \norm{\vb}^2$ (since an empty tensor product gives the trivial representation).  Note that $E^I(\vb)$ is real, since $\chi_\sigma(m)$ is real for any involution $m$.

For the variance, we will consider 
pairs of subsets $I_1, I_2 \subseteq [k]$ and decompositions of the form
\[
\vrho \otimes \vrho^*
= \left( \rho_{I_1} \otimes \rho_{I_2}^* \right) 
  \otimes \left( \rho_{\overline{I}_1} \otimes \rho_{\overline{I}_2}^* \right) 
= \left( \bigoplus_{\sigma \in \wg} 
  a^{I_1,I_2}_\sigma \sigma \right) \otimes \left( \rho_{\overline{I}_1} \otimes \rho_{\overline{I}_2}^* \right) 
\]
just as we considered $\rho \otimes \rho^*$ in the one-register case~\cite{MooreRS}.  
We then define the projection operator $\Pi^{I_1,I_2}_\sigma = \isotype^{(\rho_{I_1} \otimes \one) \otimes 
  (\rho_{I_2} \otimes \one)^*}_\sigma$ onto the subspace acted
on by $a^{I_1,I_2}_\sigma \sigma 
\otimes \rho_{\overline{I}_1} \otimes \rho_{\overline{I}_2}^*
$
and we define the following quantity, 
\begin{equation}
\label{eq:ei1i2}
E^{I_1,I_2}(\vb) \triangleq \sum_{\sigma \in \wg} 
\frac{\chi_\sigma(M)}{d_\sigma} \norm{\Pi^{I_1,I_2}_\sigma (\vb \otimes \vb^*)}^2 
\end{equation}
with $E^{\emptyset,\emptyset}(\vb)=\norm{\vb}^4$. 

We can now state the following lemma: note 
that~\eqref{eq:vark} corresponds to~\eqref{eq:var} 
in the one-register case.
\begin{lemma} \label{lem:var-k} Let $\vb \in \vrho$ and let $m$ be
  chosen uniformly from a conjugacy class $M$ of involutions.  Then
\begin{eqnarray}
  \Exp_m \norm{\Pi_m^{\otimes k} \vb}^2 
  & = & \frac{1}{2^k} \left( 1 + \sum_{I \subseteq [k] : I \neq \emptyset} 
    E^I(\vb) \label{eq:expk} \right)\enspace, \\
  \Var_m \norm{\Pi_m^{\otimes k} \vb}^2 
  & \leq & \frac{1}{4^k} \sum_{I_1,I_2 \subseteq [k] : I_1, I_2 \neq \emptyset} 
  E^{I_1,I_2}(\vb) 
  \enspace .
  \label{eq:vark}
\end{eqnarray}
\end{lemma}

\begin{proof} Let $m^I$ denote the operator that operates on the $i$th
  register by $m$ for each $i \in I$ and leaves the other registers
  unchanged.  This acts on $\vb$ as $\rho_I(m)$,
  and Lemma~\ref{lem:exp} implies that $\Exp_m \langle \vb, m^I \vb \rangle =
  E^I(\vb)$.  Then~\eqref{eq:expk} follows from the observation that
\[ \Pi_m^{\otimes k} \vb = \frac{1}{2^k} \sum_{I \subseteq [k] } m^I \vb \]
and so
\[ \Exp_m \norm{\Pi_m^{\otimes k} \vb}^2 
= \Exp_m \langle \vb, \Pi_m^{\otimes k} \vb \rangle
= \frac{1}{2^k} \sum_{I \subseteq [k] } \Exp_m \langle \vb, m^I \vb \rangle 
= \frac{1}{2^k} \sum_{I \subseteq [k] } E^I(\vb) \enspace . \]
Separating out the term $E^\emptyset(\vb) = \norm{\vb}^2$ completes the proof of~\eqref{eq:expk}.

Similarly, let the operator $m^{I_1,I_2}$ act on $\vb \otimes \vb^*$ by
multiplying the $i$th register of $\vb$ by $m$ whenever $i \in I_1$,
multiplying the $i$th register of $\vb^*$ whenever $i \in I_2$, and
leaving the other registers of $\vb$ and $\vb^*$ unchanged.  Then it
acts as $(\rho_{I_1} \otimes \rho_{I_2}^*)(m)$,
and Lemma~\ref{lem:exp} implies $\Exp_m \langle \vb \otimes \vb^*, m^{I_1,I_2}
(\vb \otimes \vb^*) \rangle = E^{I_1,I_2}(\vb)$.  Then analogous to
Lemmas~\ref{lem:second} and~\ref{lem:var}, the second moment is
\begin{align*}
  \Exp_m \norm{\Pi_m^{\otimes k} \vb}^4
  &=  \Exp_m \langle \vb, \Pi_m^{\otimes k} \vb \rangle \langle \vb^*, \Pi_m^{\otimes k} \vb^* \rangle  =  \Exp_m \langle \vb \otimes \vb^*, (\Pi_m^{\otimes k} \otimes \Pi_m^{\otimes k}) (\vb \otimes \vb^*) \rangle \\
  & = \frac{1}{4^k} \sum_{I_1,I_2 \subseteq [k] }
  \Exp_m \langle \vb \otimes \vb^*, m^{I_1,I_2} (\vb \otimes \vb^*) \rangle  = \frac{1}{4^k} \sum_{I_1,I_2 \subseteq [k] } E^{I_1,I_2}(\vb)
\end{align*}
and so the variance is
\begin{align}
  \Var_m & \norm{\Pi_m^{\otimes k} \vb}^2 
  =  \Exp_m \norm{\Pi_m^{\otimes k} \vb}^4 
  - \left( \Exp_m \norm{\Pi_m^{\otimes k} \vb}^2 \right)^2 \nonumber \\
  & =  \frac{1}{4^k} \sum_{I_1,I_2 \subseteq [k] } 
  \left( E^{I_1,I_2}(\vb) - E^{I_1}(\vb) E^{I_2}(\vb) \right) 
   =  \frac{1}{4^k} \sum_{I_1,I_2 \neq \emptyset} E^{I_1,I_2}(\vb)
  - \frac{1}{4^k} \abs{\sum_{I \neq \emptyset} E^I(\vb)}^2
  \label{eq:var-exact}
\end{align}
where we use the fact that 
$E^{I,\emptyset}(\vb) = E^{\emptyset,I}(\vb) = E^I(\vb) \norm{\vb}^2 =  E^I(\vb)$. 
Finally~\eqref{eq:vark} follows by neglecting the negative term of
\eqref{eq:var-exact}.
\end{proof}

As in the case of (one-register) Fourier sampling~\cite{MooreRS}, the
Plancherel distribution $\planch^{\otimes k}(\vrho) = d_\vrho/|G|^k$ will
play a special role in the analysis. Note that $\planch^{\otimes k}(\vrho)
= \prod \planch(\rho_i)$ and that, consistent with our conventions for
$\measured$, we elide the superscript when it will cause no confusion.

In the following two sections, we establish bounds, based on the
expressions of Lemma~\ref{lem:var-k} above, for the expectation and
variance.  Finally, we bound the expectation over $\vrho$ of the total
variation distance between the observed probability distribution
$\measured(\vrho,\vb)$ and the uniform distribution.
These bounds 
will proceed by
distinguishing a subset $\Lambda \subset \widehat{G}$ of ``bad'' representations
$\sigma$ with the undesirable property that the normalized character
$\abs{\chi_\sigma(M)/d_\sigma}$ is large; in all cases of interest, these
representations will have low dimension.

For a given $\Lambda$, we define
\[
\lambda = \lambda(M) \triangleq \max_{\sigma \notin \Lambda} \abs{\frac{\chi_\sigma(M)}{d_\sigma}}\enspace.
\]
We remark that associated with a set $\Lambda$ and a conjugacy class $M$ of
involutions one may immediately compute an upper bound on the
$\ell_1$-distance between $\measured^{\otimes k}(\cdot)$ and $\planch^{\otimes k}(\cdot)$.  
The triangle inequality and Equation~\ref{eq:prho2} imply
\begin{equation}
  \label{eq:total-variation}
  \norm{ \measured^{\otimes k} - \planch^{\otimes k}}_1 
  \leq k \norm{\measured - \planch }_1 
  \leq 2 k \bigl(\lambda + \planch(\Lambda)\bigr)
  \enspace.
\end{equation}
As we show in Section~\ref{sec:tvd-structured}, in the case relevant
to Graph Isomorphism this distance is $n^{-O(n)}$.  This allows us to
assume throughout that the $\rho_i$ are chosen according to the
Plancherel measure $\planch$ rather than to $\measured$, or
equivalently, that $\vrho$ is chosen according to the Plancherel
measure $\planch^{\otimes k}$.

\subsection{Controlling the expectation}
\label{sec:exp}

In this section we show that the expected probability distribution 
$\Exp_m \measured^{\otimes m}(\vrho,\cdot)$ is close to uniform.  
First, as we will be concerned with how representations
$\vrho$ of $G^k$ decompose into irreducible $G$-representations, we
note that for any $\sigma \in \widehat{G}$ and any $I \neq \emptyset$, 
the expected dimension of the isotypic space corresponding to $\sigma$ 
in $\rho_I \otimes \one$, namely $d_\sigma$ times the multiplicity 
$a^{\rho_I \otimes \one}_\sigma$, is given by
\begin{equation}
  \label{eq:expected-decomp}
  \Exp_{\vrho} \frac{a^{\rho_I \otimes \one}_\sigma d_\sigma}{d_\vrho} 
  = \frac{d_\sigma^2}{|G|} = \planch(\sigma) \enspace,
\end{equation}
if $\vrho$ is chosen according to the Plancherel measure~\cite{prelim:2}.  This allows us to show the following bound on the expectation of the involution projector.

\begin{lemma}
  \label{lem:boundexp}
  Let $\Lambda \subset \wg$, let $\vrho = \otimes_{i=1}^k \rho_i$ be chosen according to
  the Plancherel distribution on $\widehat{G^k}$, let $B$ be an
  arbitrary basis for $\vrho$, and let $m$ be chosen uniformly from a
  conjugacy class $M$ of involutions.  Let $\lambda=\lambda(M)$ be defined as
  above.  Then
  \[ 
  \Exp_\vrho \Exp_{\vb \in B} \abs{\Exp_m \norm{\Pi_m^{\otimes k} \vb}^2 -
    \frac{1}{2^k}} \leq \lambda + \planch(\Lambda) \enspace.
  \]
\end{lemma}

\begin{proof} For any $\vrho$ and $\vb$, Lemma~\ref{lem:var-k} and the triangle inequality 
imply that 
  \[
  \abs{\Exp_m \norm{\Pi_m^{\otimes k} \vb}^2 -
    \frac{1}{2^k}} \leq \frac{1}{2^k} \sum_{I \neq \emptyset} \sum_{\sigma \in \wg} \abs{\frac{\chi_\sigma(M)}{d_\sigma}} \norm{\Pi^I_\sigma \vb}^2
    \enspace .
  \]
  Pessimistically assuming that
  $\abs{\chi_\sigma(M)/d_\sigma} = 1$ for all $\sigma \in \Lambda$ and applying the trivial
  bound $\sum_{\sigma \notin \Lambda} \norm{\Pi_\sigma^I \vb}^2 \leq 1$ we conclude that
  \[
  \abs{\Exp_m \norm{\Pi_m^{\otimes k} \vb}^2 -
    \frac{1}{2^k}} \leq \lambda + \frac{1}{2^k} \sum_{I \neq \emptyset} \sum_{\sigma \in \Lambda} \norm{\Pi^I_\sigma \vb}^2
  \]
  Now observe that for any basis $B_\rho$ of $\vrho$ we have
  \[
  \Exp_{\vb \in B} \norm{\Pi^I_\sigma \vb}^2 = \frac{a^{\rho_I \otimes \one}_\sigma d_\sigma}{d_\vrho}
  \]
  since $a^{\rho_I \otimes \one}_\sigma d_\sigma$ is the total dimension of the isotypic subspace of $\rho_I \otimes \one$ corresponding to $\sigma$.  Applying~\eqref{eq:expected-decomp} completes the proof.
\end{proof}

\remove{
\begin{corollary} 
\label{cor:exp-simple} 
Let $\average(\vrho,\vb)$ denote $\Exp_m \measured_m^{\otimes k}(\vrho,\vb)$ and let $\uniform$ denote the uniform distribution on $B$.  Suppose that $\rank \Pi_m^{\otimes k} = d_\vrho / 2^k$ with probability 
$1-\delta$ in $\vrho$ where $\vrho$ is chosen according to the Plancherel distribution.  Then 
\[
\Exp_\vrho \norm{ \uniform - \average(\vrho,\cdot) }_{1} 
\leq 2^k \left( \lambda + \planch(\Lambda) \right) + \delta 
\enspace . 
\]
\end{corollary}
}


  
  \begin{corollary} 
    \label{cor:exp-general} 
    Let $\Lambda$ and $\lambda$ be
    defined as above and let $\vrho$ be selected according to the
    Plancherel distribution.   
    Let $\average(\vrho, \vb) = \Exp_m \measured^{\otimes k}(\vrho,\vb)$ and let
    $\uniform$ denote the uniform distribution on $B$.  Then
    \[
    \Exp_\vrho \norm{ \uniform - \average(\vrho,\cdot) }_{1} 
    \leq 2 \cdot 2^k(\lambda + \planch(\Lambda))
    \enspace.
    \]
  \end{corollary}
  
  \begin{proof}
    Define $\mathcal{I}(\vrho, \vb) = 2^k \Exp_m \norm{\Pi_m^{\otimes
        k} \vb}^2$; note that unless $\rank \Pi_m^{\otimes k} =
    d_\vrho/2^k$, this is not generally a probability distribution.
    Then Lemma~\ref{lem:boundexp} above asserts that $\Exp_\vrho
    \norm{\uniform - \mathcal{I}(\vrho,\cdot)}_1 \leq 2^k (\lambda
    +\planch(\Lambda))$.  Let $E = \{ \vrho \in \widehat{G^k} \mid
    \forall i: \rho_i \notin \Lambda \}$ and notice that as $\vrho$ is
    selected according to the Plancherel distribution, $\Pr[\vrho \in
    E] \geq 1 - k\planch(\Lambda)$. When $\vrho \in E$,
    Equation~\eqref{eq:rank} implies
    \[
    \rank \Pi_m^{\otimes k} = \frac{d_\vrho}{2^k} \prod_i \left( 1 + \frac{\chi_{\rho_i}}{d_{\rho_i}}\right) 
    \in \frac{d_\vrho}{2^k} \left[ (1 - \lambda)^k, (1 + \lambda)^k \right]
    \]
    and hence $(1 - \lambda)^k \measured(\vrho,\vb) \leq \mathcal{I}(\vrho, \vb)
    \leq (1 + \lambda)^k \measured(\vrho,\vb)$. Evidently $\|\mathcal{I}(\vrho, \cdot)
    - \measured(\vrho, \cdot)\|_1 \leq (1 + \lambda)^k - 1 \leq 2^k \lambda$.  Pessimistically assuming
    that this distance is one when $\vrho \notin E$ and using the
    triangle inequality completes the proof. 
  \end{proof}

\subsection{Controlling the variance} 
\label{sec:var}

We focus now on bounding the projectors contributing to the
$E^{I_1,I_2}$ and hence to the variance in Lemma~\ref{lem:var-k} (cf.
Equation~\eqref{eq:ei1i2}).  First, we provide a general bound on the
expectation of $\abs{\inner{\vb}{g\vb}}^2$ where $g$ ranges over the
entire group.

\begin{claim}
  \label{claim:projector-average}
  Let $\rho$ be a representation of a group $G$ acting on a space $V$ and
  let $\vb \in V$. Let $g$ be an element of $G$ chosen uniformly at
  random. Then
  \[
  \Exp_g \abs{\inner{\vb}{g\vb}}^2 \leq \sum_{\sigma \in \wg}
  \frac{\norm{\isotype_\sigma^\rho \vb}^4}{d_\sigma}\enspace.
  \]
\end{claim}

\begin{proof}
  Let $\rho \cong \oplus_j \sigma_j$, these $\sigma_j$ being irreducible, and let $V \cong \oplus
  V_j$ be the corresponding orthogonal decomposition of $V$.  Write
  $\vb = \sum_j \vb_j$ where $\vb_j \in V_j$, and $\vb_\sigma = \isotype_\sigma^\rho \vb =
  \sum_{j:\sigma_j \cong \sigma} \vb_j$. This gives
  \begin{align}
    \frac{1}{|G|} \sum_{g \in G} \abs{\inner{\vb}{g \vb}}^2 &\leq
    \frac{1}{|G|} \sum_{g \in G} \Bigl|\sum_j \inner{\vb_j}{g \vb_j}\Bigr|^2
    = \frac{1}{|G|} \sum_{g \in G} \sum_{j,k} \inner{\vb_j}{g \vb_j} \inner{\vb_k}{g \vb_k}^* \nonumber \\
    &= \sum_{j,k} \bra{\vb_j} \left(\frac{1}{|G|} \sum_{g \in G} \ket{g \vb_j}
      \bra{g \vb_k} \right) \ket{\vb_k}
    = \sum_\sigma \frac{1}{d_\sigma} \sum_{j,k: \sigma_j, \sigma_k \cong \sigma} \abs{\inner{\vb_j}{\vb_k}}^2 \label{line:schur} \\
    &\leq \sum_\sigma \frac{1}{d_\sigma} \sum_{j,k: \sigma_j, \sigma_k \cong \sigma} \norm{\vb_j}^2
    \norm{\vb_k}^2 \label{line:cs} = \sum_\sigma \frac{1}{d_\sigma} \norm{\vb_\sigma}^4
    \enspace ,
  \end{align}
  as desired.  Here we use Schur's lemma~\cite{FultonH91}
  in~\eqref{line:schur} and the Cauchy-Schwartz inequality in~\eqref{line:cs}.  
  Note that in the inner product of~\eqref{line:schur} we regard $\vb_j$ and $\vb_k$ as lying in the
  same copy of $\sigma$.
\end{proof}

\begin{lemma}
  \label{lemma:projector-bounds}
  With $\Pi^{I_1,I_2}_\sigma$ defined as in Section~\ref{sec:multi-var}, we have
\[  
\sum_{I_1,I_2} \norm{ \Pi^{I_1,I_2}_\sigma (\vb \otimes \vb^*) }^2
\leq 2^k d_\sigma^2
   \left( \sum_{I \neq \emptyset} \sum_{\tau \in \wg} \frac{\norm{\isotype_\tau^{\rho_I \otimes \one}
      \vb}^2}{d_\tau} \right)\enspace.
\]
  \remove{
  With $E^{I_1,I_2}_\sigma$ defined as in Section~\ref{sec:multi-var}, we have
  \[
  \abs{ \frac{1}{4^k} \sum_{I_1, I_2 \neq \emptyset} E^{I_1,I_2} }
   \leq \frac{1}{2^k} \left( \sum_{\sigma \in \wg} \abs{\chi_\sigma(M)} \right) 
   \left( \sum_{I \neq \emptyset} \sum_{\tau \in \wg} \frac{\norm{\isotype_\tau^{\rho_I \otimes \one}
      \vb}^2}{d_\tau} \right) 
      \enspace.
  \]
  }
\end{lemma}

\begin{proof}
  We can write $\Pi^{I_1,I_2}_\sigma$ as $\isotype^{(\rho_{I_1}
    \otimes \one) \otimes (\rho_{I_2} \otimes \one)^*}_\sigma$, where
  $\one$ and $\one^*$ act on $\rho_{\overline{I}_1}$ and
  $\rho_{\overline{I}_2}^*$ respectively.  Using the same notation as
  in Section~\ref{sec:multi-var}, let $g^{I_1,I_2}$ act on $\vb
  \otimes \vb^*$ by multiplying the $i$th register of $\vb$ by $m$
  whenever $i \in I_1$, multiplying the $i$th register of $\vb^*$
  whenever $i \in I_2$, and leaving the other registers of $\vb$ and
  $\vb^*$ unchanged. From the defining expression of Definition~\ref{def:isotypic} we have
  \[ 
  \norm{ \Pi^{I_1,I_2}_\sigma (\vb \otimes \vb^*) }^2
  = \frac{d_\sigma}{|G|} \sum_{g \in G} \chi_\sigma(g)^* 
  \inner{\vb \otimes \vb^*}{g^{I_1,I_2} (\vb \otimes \vb^*)} 
  = \frac{d_\sigma}{|G|} \sum_{g \in G} \chi_\sigma(g)^* 
  \inner{\vb}{g^{I_1}\vb}\inner{\vb}{g^{I_2}\vb}^*
  \enspace.
  \]
  Observe, however, that
  \begin{align*}  
  \sum_{I_1,I_2} & \norm{ \Pi^{I_1,I_2}_\sigma (\vb \otimes \vb^*) }^2
  = \frac{d_\sigma}{|G|} \sum_{g \in G} \chi_\sigma(g)^* \sum_{I_1,I_2}
  \inner{\vb}{g^{I_1}\vb}\inner{\vb}{g^{I_2}\vb}^* 
  = \frac{d_\sigma}{|G|} \sum_{g \in G} \chi_\sigma(g)^* 
  \abs{ \sum_I \inner{\vb}{g^I \vb}}^2 \\
&  \leq \frac{d_\sigma^2}{|G|} \sum_{g \in G} 
  \abs{ \sum_I \inner{\vb}{g^I \vb}}^2 
  \leq 2^k \frac{d_\sigma^2}{|G|} \sum_{g \in G} 
  \sum_I \abs{\inner{\vb}{g^I \vb}}^2 
  = 2^k d_\sigma^2 \sum_I \Exp_g \abs{\inner{\vb}{g^I \vb}}^2 
  \end{align*}
  by the triangle inequality and Cauchy-Schwarz.  Finally, we apply
  Claim~\ref{claim:projector-average} to the expectations above and
  use the fact that as $\norm{\vb} = 1$, $\norm{\Pi \vb}^4 \leq \norm{\Pi
    \vb}^2$ for any projection operator $\Pi$.  \remove{ The statement
    of the lemma then follows by applying the triangle inequality to
    the definition of $E^{I_1,I_2}$,
  \begin{align*}
  \abs{ \sum_{I_1, I_2 \neq \emptyset} E^{I_1,I_2} } 
  &\leq \sum_{I_1, I_2 \neq \emptyset} \abs{E^{I_1,I_2}} 
  \leq \sum_{\sigma \in \wg} \abs{\frac{\chi_\sigma(M)}{d_\sigma}}
  \sum_{I_1,I_2} \norm{ \Pi^{I_1,I_2}_\sigma (\vb \otimes \vb^*) }^2
  \enspace . 
  \end{align*}
  }
\end{proof}

\noindent
Then the following lemma bounds the variance of $\norm{\Pi_m^{\otimes k} \vb}^2$ just as Lemma~\ref{lem:boundexp} bounds the expectation.

\begin{lemma} 
\label{lem:boundvar} 
Let $\Lambda \subset \wg$, let $\vrho =
  \otimes_{i=1}^k \rho_i$ where the $\rho_i$ are independently chosen according to
  the Plancherel distribution, let $B$ be an arbitrary basis for
  $\vrho$, and let $m$ be chosen uniformly from a conjugacy class $M$ of
  involutions.  Let $\lambda=\lambda(M)$ be defined as above.  Then
  \[ 
  \Exp_\vrho \Exp_{\vb \in B} \Var_m \norm{\Pi_m^{\otimes k} \vb}^2 
  \leq \Delta \triangleq \lambda + \planch(\Lambda) \left( \sum_{\tau \in \wg} d_\tau \right) 
  \enspace .
  \]
\end{lemma}

\begin{proof}
  Applying Lemma~\ref{lemma:projector-bounds} to control the terms in
  $E^{I_1,I_2}(\vb)$ where $\sigma \in \Lambda$, pessimistically
  assuming that $\abs{\chi_\sigma(M)/d_\sigma} = 1$ for all $\sigma
  \in \Lambda$, and using the obvious bound $\sum_{\sigma \notin
    \Lambda} \norm{\Pi^{I_1,I_2}_\sigma(\vb \otimes \vb^*)}^2 \leq 1$
  for the others, we see from~\eqref{eq:vark} that
  \begin{align*}
    \Var_m & \norm{\Pi_m^{\otimes k} \vb}^2 \leq \frac{1}{4^k}
    \sum_{I_1,I_2 \neq \emptyset} E^{I_1,I_2}(\vb) \leq \frac{1}{4^k}
    \sum_{I_1,I_2} \sum_{\sigma \in \wg}
    \abs{\frac{\chi_\sigma(M)}{d_\sigma}} \norm{\Pi_\sigma^{I_1,I_2} (\vb \otimes \vb^*)}^2 \\
    &\leq \lambda + \frac{1}{4^k} \sum_{I_1,I_2} \sum_{\sigma \in
      \Lambda} \norm{\Pi_\sigma^{I_1,I_2} \vb \otimes \vb^*}^2 =
    \lambda + \frac{1}{2^k} \left(\sum_{\sigma \in \Lambda} d_\sigma^2
    \right) \sum_{I \neq \emptyset} \sum_{\tau \in \wg}
    \frac{\norm{\isotype_\tau^{\rho_I \otimes \one} \vb}^2}{d_\tau}
    \enspace.
  \end{align*}
  Now we take the expectation of this over the basis $B$.  Since
  $\Exp_{\vb \in B} \norm{\isotype_\tau^{\rho_I \otimes \one} \vb}^2 =
  a^{\rho_I \otimes \one}_\tau d_\tau/d_\vrho$, we have
  \[
  \Exp_{\vb \in B} \Var_m \norm{\Pi_m^{\otimes k} \vb}^2 \leq \lambda
  + \frac{1}{2^k} \left( \sum_{\sigma \in \Lambda} d_\sigma^2\right)
  \sum_{I \neq \emptyset} \sum_{\tau \in \wg} \frac{a^{\rho_I \otimes
      \one}_\tau}{d_\vrho}
  \]
  and Equation~\eqref{eq:expected-decomp} completes the proof.
\end{proof}

\subsection{Bounding the total variation distance}
\label{sec:tvd}

Finally, the next lemma relates the bound of Lemma~\ref{lem:boundvar}
to the expected variation distance of the observed distribution from
the uniform distribution.
\begin{lemma}  
\label{lem:tvd}
    Let $\Lambda$ and $\lambda$ be
    defined as above, let $\vrho$ be selected according to the
    Plancherel distribution, and let $m$ be uniformly random in its conjugacy class. 
    Let $B$ be a basis for $\vrho$ and let
    $\uniform$ denote the uniform distribution on $B$.  Then
\begin{gather*}
  \Exp_\vrho \Exp_m \norm{\measured(\vrho,\cdot) - \uniform}_1
  \leq 2^k \left[ (1-\lambda)^{-k} \sqrt{\Delta} + 3 \cdot (\lambda + \planch(\Lambda)) \right] 
  \remove{
 \; \mbox{where } \;
  \Delta = \sqrt{ \lambda + \frac{1}{|G|} \Bigl( \sum_{\sigma \in \Lambda} d_\sigma
    \Bigr) \Bigl( \sum_{\tau \in \wg} d_\tau \Bigr) } 
  \enspace .
  }
\end{gather*}
where $\Delta$ is defined as in Lemma~\ref{lem:boundvar}.
\end{lemma}

\begin{proof}
  As in Corollary~\ref{cor:exp-general}, let $\average(\vrho,\vb)$ denote $\Exp_m
  \measured(\vrho,\vb)$.  Then we have, analogous to Lemma~\ref{cor:exp-general}, 
  \begin{align*}
    \Exp_\vrho &\Exp_m \norm{\measured(\vrho,\cdot) - \average(\vrho,\cdot)}_1
    = \Exp_\vrho \Exp_m \sum_{\vb \in B} \abs{\measured(\vrho,\vb) - \average(\vrho,\vb)} \\
    &\leq \Exp_\vrho \Exp_m \sqrt{d_\vrho^2 \Exp_{\vb \in B}
      \abs{\measured(\vrho,\vb) - \average(\vrho,\vb)}^2} \leq \Exp_\vrho
    \sqrt{\Exp_m d_\vrho^2 \Exp_{\vb \in B}
      \abs{\measured(\vrho,\vb) - \average(\vrho,\vb)}^2} \\
    &= \Exp_\vrho \sqrt{\Exp_{\vb \in B} d_\vrho^2 \Var_{m}
      \measured(\vrho,\vb)} \leq 2^k (1-\lambda)^{-k} \Exp_\vrho \sqrt{\Exp_{\vb \in B} \Var_{m} \norm{\Pi_m^{\otimes k} \vb}^2 } + k \planch(\Lambda) \\
    &\leq 2^k (1-\lambda)^k \sqrt{ \Exp_\vrho \Exp_{\vb \in B} \Var_{m} \norm{\Pi_m^{\otimes k}
        \vb}^2 } + k \planch(\Lambda) \enspace .
  \end{align*}
  The proof is completed by Lemma~\ref{lem:boundvar}, Corollary~\ref{cor:exp-general}, and the triangle inequality.
\end{proof}

\section{The total variation distance}
\label{sec:tvd-structured}

Having established the generic bounds of the previous sections, it remains 
simply to apply them to a given group, using a description of its irreducible 
representations and a choice of the subset $\Lambda$.  The standard 
reduction from Graph Isomorphism yields permutations of $2n$ objects, 
namely the vertices of two graphs of $n$ vertices each.  However, rather than 
all of $S_{2n}$, it suffices to consider the subgroup $K=S_n \wr \Z_2 \subset S_{2n}$  
consisting of permutations which either fix the two vertex sets or swap them.  

The irreducible representations of $K$ and their characters are discussed in the 
Appendix.  Our choice of ``bad'' representations $\Lambda \subset \wk$ 
consists of those induced up from representations $\rho \otimes \rho$ of 
$S_n \times S_n$ with the property that $d_\rho < n^{n/5}$.  Simple counting arguments 
then show that $\lambda \leq n^{-n/5}$, $\planch(\Lambda) = n^{-6n/5} e^{O(n)}$, 
and $\Delta = n^{-n/5} e^{O(n)}$ where $\Delta$ is as defined in Lemma~\ref{lem:boundvar}.  
With the understanding that $k = n^{O(1)}$, we have $(1 - \lambda)^{-k} = 1 + o(1)$ 
and we find that the expected variation distance in Lemma~\ref{lem:tvd} is
\[
\Exp_\vrho \Exp_m \norm{\measured(\vrho,\cdot) - \uniform}_1 \leq 2^k
n^{-n/10} e^{O(n)} \enspace .
\]
Thus if $k < C n \log_2 n$ where $C$ is bounded below $1/10$, 
this is $n^{-\Omega(n)}$, and by Markov's inequality the probability in
$\vrho$ and $m$ that $\norm{\measured(\vrho,\cdot) - \uniform}_1 >
n^{-\Omega(n)}$ is no more than $n^{-\Omega(n)}$.  Finally, since 
by Equation~\ref{eq:total-variation} $\norm{\measured(\cdot)-\planch(\cdot)}_1
\leq 2 (\lambda + \planch(\Lambda)) = n^{-\Omega(n)}$, any event that holds
with probability $Q$ in $\planch(\cdot)$ holds with probability
$Q-n^{-\Omega(n)}$ in $\measured(\cdot)$.  This completes the proof of
Theorem~\ref{thm:main}; we have made no effort to optimize the constant $C$.

\remove{
Because we wished here simply to show that a superpolynomial number of
experiments are necessary unless $k = \Omega(n \log n)$, our analysis
assumes the worst whenever a register is found to be in one of the
representations $\tau_{\{\rho,\rho\},\one}$ or $\tau_{\{\rho,\rho\},\pi}$ where the character
at $M$ is nonzero.  However, the normalized characters of almost all
of these representations, namely $1/d_\rho$, is of order $n^{-\Omega(n)}$:
therefore, the observed distribution is in fact exponentially close to
the Plancherel distribution, and $\rank \Pi_m^{\otimes k} / d_\vrho$ is
exponentially close to $1/2^k$ with probability exponentially close to
$1$.  Thus we claim that the terms $e^{-\Omega(\sqrt{n})}$ in
Theorem~\ref{thm:main} can be improved to $n^{-\Omega(n)}$.
}

We remark that these bounds can be established if $S_n$ is replaced
with any group $G$ for which a sufficient fraction of the Plancherel
measure lies on high-dimensional representations.  For any such group,
the hidden subgroup problem on $G \wr \Z_2$ requires entangled
measurements on $\Theta(n \log n)$ coset states.

\section*{Acknowledgments.}  
This work was supported by the NSF under grants EIA-0218443,
EIA-0218563, CCR-0220070, CCR-0220264, and CCF-0524613, and the ARO
under grant W911NF-04-R-0009.  We are grateful to Sean Hallgren for
informing us of his work with Martin R\"otteler and Pranab Sen.  We thank Tracy Conrad and Sally Milius for their support
and tolerance.  C.M.\ also thanks Rosemary Moore for providing a
larger perspective.

\appendix

\section{The group generated by structured involutions}
\label{sec:representation-theory}

In this section we review the representation theory of the symmetric group $S_n$, and describe the representations of the subgroup of $S_{2n}$ relevant to Graph Isomorphism.  First, recall that the irreducible representations $\rho$ of $S_n$ are labeled by Young diagrams, or equivalently integer partitions $\lambda_1 \geq \lambda_2 \geq \cdots \geq \lambda_t$ such that $\sum_i \lambda_i = n$.  The number of irreducible representations is then the partition number $p(n)=e^{O(\sqrt{n})}$.

In the standard reduction from Graph Isomorphism, we consider
subgroups $\{1,m\}$ where $m$ is an involution consisting of $n$
disjoint transpositions, matching each vertex in one graph with the
corresponding vertex in the other.  However, rather than considering
all such conjugates in $S_{2n}$, it makes sense to focus on those
involutions $m$ which map $\{1,\ldots,n\}$ to $\{n+1,\ldots,2n\}$, which we identify
with the vertex sets $V_1$ and $V_2$ of the two graphs.  Such $m$ lie
inside a subgroup of $S_{2n}$: namely, if $s$ denotes a canonical
involution $(1\;n+1) (2\; n+2) \ldots (n\; 2n)$, then $m=\alpha^{-1} s \alpha$
where $\alpha$ permutes $V_1$.

The set of all such involutions generates a subgroup $K$ of $S_{2n}$.
Let $S_{n,n}$ denote the {\em Young subgroup} $S_{n,n}$ which fixes
the sets $V_1$ and $V_2$; then $K$ is the subgroup generated by
$S_{n,n}$ and $s$.  Algebraically, $K$ is the \emph{wreath product}
$S_n \wr \Z_2$, and can also be written as a semidirect product $K =
(S_n \times S_n) \rtimes \Z_2$.  If $\alpha, \beta \in S_n$ and $t \in \Z_2$, we denote
by $((\alpha,\beta),t)$ the element which applies $\alpha$ to $V_1$ and $\beta$ to
$V_2$, and then applies $s^t$.  Note that $|K|=2 n!^2=n^{2n}
e^{-O(n)}$.

\begin{sloppypar}
We can determine $K$'s irreducible representations and their characters as follows.  For two irreducible
representations $\rho$ and $\sigma$ of $S_n$, let $\rho \boxtimes \sigma$ denote their
tensor product as a representation of $S_{n,n} \cong S_n \times S_n$.  We
consider the induced representation $\trs = \Ind_{S_{n,n}}^K (\rho \boxtimes
\sigma)$ and denote its character $\chi_{\{\rho,\sigma\}}$.  It is easy to see that
$$
\chi_{\{\rho,\sigma\}}\bigl( ((\alpha, \beta),t) \bigr) = \begin{cases} 0 & \text{if}\;t=1\\
  \chi_\rho(\alpha)\chi_\sigma(\beta) +
  \chi_\sigma(\alpha)\chi_\rho(\beta) & \text{if}\;t=0\enspace;
\end{cases}
$$
as the notation suggests, this depends only on the multiset $\{ \rho, \sigma\}$.
An easy computation shows that $\langle \chi_{\{\rho,\sigma\}}, \chi_{\{\rho, \sigma\}}\rangle = 1+\delta_{\rho,\sigma}$.  Thus, if $\rho \not\cong \sigma$, then $\trs$ is irreducible of dimension $2 d_\rho d_\sigma$; while if $\rho \cong \sigma$ then it decomposes into two irreducible representations of dimension $d_\rho^2$, 
\[ \tau_{\{\rho,\rho\} }   
\cong 
\tau_{\{\rho,\rho\},\one} \oplus \tau_{\{\rho,\rho\},\pi} 
\]
where $\one$ and $\pi$ are the trivial and sign representations, respectively, of $\Z_2$.
Each of these irreducible representations acts on $V_\rho \otimes
V_\rho$, the vector space supporting the action of $\rho \boxtimes \rho$.  
Both of them realize the element $((\alpha,\beta),0)$ as the
linear map $\rho(\alpha) \otimes \rho(\beta)$, while $\tau_{\{\rho,\rho\},\one}$ 
and $\tau_{\{\rho,\rho\},\pi}$ 
realize the element $((1,1),1)$ as the maps which send $\vec{u} \otimes \vec{v}$ 
to $\vec{v} \otimes \vec{u}$ and $-\vec{v} \otimes \vec{u}$ respectively. 
The characters of these representations are
$$
\chi_{\{\rho,\rho\},\one} ((\alpha, \beta),t) = \begin{cases} \chi_\rho(\alpha) + \chi_\rho(\beta) & \text{if}\;t = 0 \\
  \chi_\rho(\alpha\beta) & \text{if}\;t=1 \end{cases} \enspace , 
\qquad
\chi_{\{\rho,\rho\},\pi} ((\alpha, \beta),t) = \begin{cases} \chi_\rho(\alpha) + \chi_\rho(\beta) & \text{if}\;t = 0 \\
  -\chi_\rho(\alpha\beta) & \text{if}\;t=1 \end{cases} 
  \enspace . 
$$
In particular, since $m$ is of the form $((\alpha,\alpha^{-1}),1)$, we have the normalized characters
\begin{align}
\label{eq:outside}
\frac{\chi_{\{\rho,\rho\},\one}(m)}{d_{\{\rho,\rho\},\one}} = \frac{1}{d_\rho} \, , \; \;
\frac{\chi_{\{\rho,\rho\},\pi}(m)}{d_{\{\rho,\rho\},\pi}} = -\frac{1}{d_\rho} 
\end{align}
and $\chi_{\{\rho,\sigma\}}(m) = 0$ for all $\rho \not\cong \sigma$.
\end{sloppypar}

We remark that this construction of the irreducible representations
and their characters works for any group of the form $G \wr \Z_2$. In
particular, the normalized characters of the involutions that ``swap''
the two copies of $G$ are either $0$ or $\pm 1/d_\rho$ for some $\rho \in
\widehat{G}$.

If we choose $\Lambda$ to consist of those $\tau_{\{\rho,\rho\},\one}$ 
and $\tau_{\{\rho,\rho\},\pi}$ such that $d_\rho < n^{n/5}$, then by 
by~\eqref{eq:outside} we have $\lambda \leq n^{-n/5}$.  
Since there are at most $p(n)^2$ irreducible representations of $K$ we have 
\[ 
\planch(\Lambda) = \sum_{\tau \in \Lambda} d_\tau^2/|K| 
\leq p(n)^2 \,n^{4n/5} / |K| = n^{-6n/5} e^{O(n)}
\enspace . 
\]
Similarly, since no irreducible representations of $K$ can have dimension greater than $\sqrt{|K|}$, 
the quantity $\Delta$ defined in Lemma~\ref{lem:boundvar} is bound by
\[ \Delta = \lambda + \frac{\planch(\Lambda)}{|K|} \left( \sum_{\tau \in \wg} d_\tau \right)  
  \leq n^{-n/5} +  n^{-6n/5}\,p(n)^4\,\sqrt{|K|}\,e^{O(n)}
  = n^{-n/5} e^{O(n)} \enspace . 
\]


\begin{thebibliography}{99}

\newcommand{\proc}{Proc.\ }

\bibitem{BaconCvD1} David Bacon, Andrew Childs, and Wim van Dam.
\newblock Optimal measurements for the dihedral hidden subgroup problem.
\newblock Preprint, quant-ph/0501044 (2005).

\bibitem{BaconCvD2} David Bacon, Andrew Childs, and Wim van Dam.
  \newblock From optimal measurement to efficient quantum algorithms for the hidden subgroup problem over semidirect product groups.
  \newblock \emph{\proc 46th Symposium on Foundations of Computer Science},
  2005.

\bibitem{Beals97} Robert Beals.  
\newblock Quantum computation of Fourier transforms over symmetric groups.  
\newblock \emph{\proc 29th Annual {ACM} Symposium
on the Theory of Computing}, pages 48--53, 1997.

\bibitem{BernsteinV93}
Ethan Bernstein and Umesh Vazirani.
\newblock Quantum complexity theory (preliminary abstract).
\newblock \emph{\proc 25th Annual ACM Symposium on the
  Theory of Computing}, pages 11--20, 1993.

\bibitem{ChildsW}
Andrew Childs and Pawe{\l} Wojcan, 
\newblock On the quantum hardness of solving isomorphism problems as nonabelian hidden shift problems.
\newblock Preprint, quant-ph/0510185 (2005).

\bibitem{vanDamHI03}
Wim van Dam, Sean Hallgren, and Lawrence Ip. 
\newblock Quantum algorithms for some hidden shift problems. 
\newblock \emph{\proc 14th ACM-SIAM Symposium on Discrete Algorithms},   
pages 489--498, 2003.
 
\bibitem{EttingerH98}
Mark Ettinger and Peter H{\o}yer.
\newblock On quantum algorithms for noncommutative hidden subgroups.
\newblock Preprint, quant-ph/9807029 (1998).

\bibitem{EttingerHK04}
Mark Ettinger and Peter H{\o}yer and Emmanuel Knill.
\newblock The quantum query complexity of the hidden subgroup problem is 
polynomial. 
\newblock \emph{Information Processing Letters}, to appear.

\bibitem{FriedlIMSS02}
Katalin Friedl, {G{\'a}bor} Ivanyos, {Fr{\'e}d{\'e}ric} Magniez, Miklos Santha,
  and Pranab Sen.
\newblock Hidden translation and orbit coset in quantum computing.
\newblock \emph{\proc 35th ACM Symposium on Theory of
Computing}, 2003.

\bibitem{FultonH91}
William Fulton and Joe Harris.
\newblock \emph{Representation Theory: A First Course}.
\newblock Number 129 in Graduate Texts in Mathematics. Springer-Verlag, 1991.

\bibitem{GrigniSVV01}
Michelangelo Grigni, Leonard~J. Schulman, Monica Vazirani, and Umesh Vazirani.
\newblock Quantum mechanical algorithms for the nonabelian hidden subgroup
  problem.
\newblock \emph{\proc 33rd ACM Symposium on Theory of
  Computing}, pages 68--74, 2001.
  
\bibitem{HalesH99}
Lisa Hales and Sean Hallgren.
\newblock Quantum Fourier sampling simplified.
\newblock \emph{\proc 31st Annual ACM Symposium on Theory of Computing}, 1999.

\bibitem{HalesH00}
Lisa Hales and Sean Hallgren.
\newblock An improved quantum Fourier transform algorithm and applications.
\newblock \emph{\proc 41st Annual Symposium on Foundations of Computer Science},  
2000.

\bibitem{HallgrenRT00}
Sean Hallgren, Alexander Russell, and Amnon {Ta-Shma}.
\newblock Normal subgroup reconstruction and quantum computation using group
  representations.
\newblock \emph{\proc 32nd ACM Symposium on Theory of
  Computing}, pages 627--635, 2000.

\bibitem{Hoyer97}
Peter H{\o}yer.
\newblock Efficient quantum transforms.
\newblock Preprint, quant-ph/9702028 (1997).

\bibitem{InuiLeGall}
Yoshifumi Inui and Fran\c{c}ois Le Gall.
\newblock An efficient algorithm for the hidden subgroup problem over a class of semi-direct product groups.
\newblock \proc EQIS 2004.

\bibitem{Ip}
Lawrence Ip.
\newblock Shor's algorithm is optimal.
\newblock Preprint, 2004.

\bibitem{IvanyosMS01}
G{\'a}bor Ivanyos, Fr{\'e}d{\'e}ric Magniez, and Miklos Santha.
\newblock Efficient quantum algorithms for some instances of the non-{abelian}
  hidden subgroup problem.
\newblock \emph{Int. J. Found. Comput. Sci.} 14(5): 723--740, 2003.

\bibitem{Jozsa00}
Richard Jozsa.
\newblock Quantum factoring, discrete logarithms and the hidden subgroup
  problem.
\newblock Preprint, quant-ph/0012084 (2000).

\bibitem{KempeS} Julia Kempe and Aner Shalev.
\newblock The hidden subgroup problem and permutation group theory.
\newblock Preprint, quant-ph/0406046 (2004).

\bibitem{Kerov} {S. V.} Kerov.
  \newblock \emph{Asymptotic
    representation theory of the symmetric group and its applications
    in analysis}.
  \newblock Translated by {N. V.} Tsilevich.
  \newblock Volume 219 in Translations of Mathematical Monographs. American
  Mathematical Society, 2003.

\bibitem{Koiran05}
Pascal Koiran, Vincent Nesme, and Natacha Portier.
\newblock A quantum lower bound for the query complexity of {Simon's}
problem.
\newblock \emph{\proc of the 32nd
International Colloquium on
Automata, Languages and Programming}, 2005.

\bibitem{Kuperberg03}
Greg Kuperberg.
\newblock A subexponential-time quantum algorithm for the dihedral hidden 
subgroup problem. 
\newblock Preprint, quant-ph/0302112 (2003).

\bibitem{MooreRR04}
Cristopher Moore, Daniel Rockmore, and Alexander Russell.
\newblock Generic quantum Fourier transforms. 
\newblock \emph{\proc 15th Annual ACM-SIAM Symposium on Discrete Algorithms}, 
pages 778--787, 2004.

\bibitem{MooreRRS04} Cristopher Moore, Daniel Rockmore, Alexander Russell, and Leonard Schulman.
\newblock The value of basis selection in Fourier sampling: hidden subgroup problems for affine groups.
\newblock \emph{\proc 15th Annual ACM-SIAM Symposium on Discrete Algorithms}, 
pages 1113--1122, 2004.

\bibitem{MooreR:banff} Cristopher Moore and Alexander Russell.
\newblock Explicit multiregister measurements for hidden subgroup problems; or, Fourier sampling strikes back.
\newblock Preprint, quant-ph/0504067 (2005).

\bibitem{MooreR:gelfand} Cristopher Moore and Alexander Russell.
\newblock For distinguishing conjugate hidden subgroups, the pretty good measurement is as good as it gets.
\newblock Preprint, quant-ph/0501177 (2005).

\bibitem{MooreRS} Cristopher Moore and Alexander Russell and Leonard
  Schulman. 
  \newblock The symmetric group defies Fourier sampling.
  \newblock \emph{\proc 46th Symposium on Foundations of Computer Science}, pages 479--488
  (2005).

\bibitem{prelim:2} Cristopher Moore and Alexander Russell.
\newblock The symmetric group defies strong Fourier sampling: part II.
\newblock Preprint, quant-ph/0501066 (2005).

\bibitem{prelim:multi} Cristopher Moore and Alexander Russell.
\newblock Quantum Measurements for Graph Isomorphism Require Entanglement: Tight Results on Multiregister Fourier Sampling
\newblock Preprint, quant-ph/0510233 (2005).

\bibitem{RadhakrishnanRS}
Jaikumar Radhakrishnan, Martin R\"{o}tteler, and Pranab Sen.
\newblock On the Power of Random Bases in Fourier Sampling: Hidden Subgroup Problem in the Heisenberg Groups.
\newblock \emph{\proc 32nd International Colloquium on
Automata, Languages and Programming} (2005).

\bibitem{Regev} Oded Regev. 
\newblock Quantum computation and lattice problems.  
\newblock \emph{\proc 43rd Symposium on Foundations of Computer Science}, 
pages 520--530, 2002.

\bibitem{RoettelerB98}
Martin R\"{o}tteler and Thomas Beth.
\newblock Polynomial-time solution to the hidden subgroup problem for a class
  of non-abelian groups.
\newblock Preprint, quant-ph/9812070 (1998).

\remove{
\bibitem{Roichman96}
Yuval Roichman.
\newblock Upper bound on the characters of the symmetric groups.
\newblock \emph{Inventiones Mathematicae}, 125:451--485, 1996. 
}

\remove{
\bibitem{Roman92}
Steven Roman.
\newblock \emph{Advanced Linear Algebra}.
\newblock Number 135 in Graduate Texts in Mathematics. Springer, 1992.
}

\bibitem{Serre77}
Jean-Pierre Serre.
\newblock \emph{Linear Representations of Finite Groups}.
\newblock Number~42 in Graduate Texts in Mathematics. Springer-Verlag, 1977.

\bibitem{Shor97}
Peter~W. Shor.
\newblock Polynomial-time algorithms for prime factorization and discrete
  logarithms on a quantum computer.
\newblock \emph{{SIAM} Journal on Computing}, 26(5):1484--1509, 1997.

\bibitem{Simon97}
Daniel~R. Simon.
\newblock On the power of quantum computation.
\newblock \emph{{SIAM} Journal on Computing}, 26(5):1474--1483, 1997.

\bibitem{VershikK} {A. M.} Vershik and {S. V.} Kerov.  \newblock
  Asymptotic behavior of the maximum and generic dimensions of
  irreducible representations of the symmetric group.  \newblock Funk.
  Anal. i Prolizhen, 19(1):25--36, 1985; \newblock English
  translation, Funct. Anal. Appl., 19:21--31, 1989.
  
\end{thebibliography}
\end{document}